\documentclass[aps,prd,onecolumn,groupedaddress,showpacs,nofootinbib,amssymb]{revtex4}
\usepackage[dvips]{graphicx}
\usepackage{amssymb}
\usepackage{amsmath}
\usepackage{graphicx,,color}
\usepackage{amsfonts}
\usepackage{bm}
\usepackage{cancel}
\usepackage{comment}
\newcommand\be{\begin{equation}}
\newcommand\ee{\end{equation}}

\allowdisplaybreaks[4]

\begin{document}

\tolerance=5000

\title{On the Occurrence of Finite-time Singularities in Swampland-related Quintessence Dark Energy Models}
\author{V.K.~Oikonomou,$^{1,2}$\,\thanks{v.k.oikonomou1979@gmail.com}}
\author{Achilles Gitsis,$^{1}$\,\thanks{agitsis62@gmail.com}}
\author{Maria Mitrou,$^{1}$\thanks{mmitrou@physics.auth.gr}}
\affiliation{$^{1)}$ Department of Physics, Aristotle University
of Thessaloniki, Thessaloniki 54124, Greece
\\
$^{2)}$ Laboratory for Theoretical Cosmology, Tomsk State
University of Control Systems and Radioelectronics, 634050 Tomsk,
Russia (TUSUR)}

\tolerance=5000

\begin{abstract}
In this work we focus on the phase space singularities of interactive
quintessence model in the presence of matter fluid. This model is
related to swampland studies, that the outcomes affect all these
Swampland related models with the same dynamical system. We shall
form the dynamical system corresponding to the cosmological
system, which is eventually autonomous, and by using the dominant
balances technique we shall investigate the occurrence or not of
finite-time singularities. Our results indicate that the dynamical
system of the model may develop finite-time singularities, but
these are not general singularities, like in the case that the
matter fluids were absent, in which case singularities occurred
for general initial conditions. Hence, the presence of matter
fluids affects the dynamical system of the cosmological system,
making the singularities to depend on the initial conditions,
instead of occurring for general initial conditions.
\end{abstract}

\maketitle

\section{Introduction}

Undoubtedly string theory in its various M-theory forms seems to
be the perfect candidate for describing the UV-completion of the
Standard Model of particle physics and gravity. However the
predictions of string theory are highly unlikely to be ever
observed experimentally in terrestrial experiments. Therefore,
although a beautiful theory, it seems that for the time being
remains just a candidate UV-completion theory. Nevertheless,
string theory may constrain the high energy limits of the
classical theory. One such example are the Swampland criteria,
firstly introduced in Refs. \cite{Vafa:2005ui,Ooguri:2006in} and
were further developed in Refs.
\cite{Palti:2020qlc,Mizuno:2019bxy,Brandenberger:2020oav,Blumenhagen:2019vgj,Wang:2019eym,Wang:2018duq,Benetti:2019smr,Palti:2019pca,Cai:2018ebs,Akrami:2018ylq,Mizuno:2019pcm,Aragam:2019khr,Brahma:2019mdd,Mukhopadhyay:2019cai,Brahma:2019kch,Haque:2019prw,Heckman:2019dsj,Acharya:2018deu,Elizalde:2018dvw,Cheong:2018udx,Heckman:2018mxl,Kinney:2018nny,Garg:2018reu,Lin:2018rnx,Park:2018fuj,Olguin-Tejo:2018pfq,Fukuda:2018haz,Wang:2018kly,Ooguri:2018wrx,Matsui:2018xwa,Obied:2018sgi,Agrawal:2018own,Murayama:2018lie,Marsh:2018kub,Storm:2020gtv,Trivedi:2020wxf,Sharma:2020wba,Odintsov:2020zkl,Mohammadi:2020twg,Trivedi:2020xlh,Han:2018yrk,Achucarro:2018vey,Akrami:2020zfz,Odintsov:2018zai,Oikonomou:2021zfl,Oikonomou:2020oex},
see also \cite{Colgain:2018wgk,Colgain:2019joh,Banerjee:2020xcn}
for the Swampland criteria implications on the $H_0$ tension
problem. The Swampland criteria basically constrain several
scalar field parameters at low-energies compared to the mother
M-theory. In a previous work \cite{Odintsov:2018zai} we studied
the singularity structure of the dynamical system corresponding to
scalar field theories used for the Swampland models in vacuum. In
this paper we extend the work \cite{Odintsov:2018zai} to include
perfect matter fluids, and specifically dark matter and radiation
fluids. The corresponding dynamical system is perfectly studied in
\cite{Wang:2018duq}. Our aim is to see whether the matter fluids
affect the finite-time singularity structure of the phase space of
the scalar field model. As we show, indeed the matter fluids
affect the finite-time structure of the phase space of the model,
and in the present case scenario, singularities typically exist,
but only for limited sets of initial conditions in the phase
space. This is in contrast with the results obtained in our
previous work \cite{Odintsov:2018zai}, where finite-time
singularities occurred for a general set of initial conditions.

In the present work, we use natural units $c = \hbar = 1$ and the
metric at use is the flat Friedmann Robertson Walker (FRW) metric
whose line element is given by
\begin{equation}
    \centering
    \label{frw}
    d s^2 = - d t^2 + a(t) \sum_{i = 1}^3 d x_i^2,
\end{equation}
where $a (t)$ is the scale factor. The Ricci scalar for the FRW metric is
\begin{equation}
    \centering
    \label{ricci}
    R = 6\dot{H} + 12 H^2,
\end{equation}
where $H=\frac{\dot{a}}{a}$ is the Hubble parameter. We also adopt reduced Planck units, that is $\kappa =\frac{1}{M_p}= 1$.

\section{Essential Features of Dominant Balances Methodology}

Our aim is to study the finite-time singularities structure of the
dynamical system developed in \cite{Wang:2018duq}. To this end we
shall use the method of dominant balances, see Refs.
\cite{Odintsov:2018zai} and \cite{Odintsov:2018uaw} and references
therein for more details. We brief it here for reasons of
completeness.
\begin{itemize}
    \item Consider a dynamical system of $n$ differential equations of the form
    \begin{equation}
        \centering
        \label{dym}
        \dot{x}_i = f_i(x),
    \end{equation}
where $i  =1, 2, \dots n$. Approaching the region of the
singularity, we can extract from $f_i$ the part that becomes
considerable, we shall call it from now on \textit{dominant part}.
This dominant part constitutes a mathematically consistent
truncation of the system and denote it as $\hat{f}_i$. Now,
(\ref{dym}) has become
\begin{equation}
        \centering
        \label{sys}
        \dot{x}_i = \hat{f}_i(x).
    \end{equation}
We should note that the dot denotes differentiation with respect
to time $t$. In our case, in spite of $t$ the e-foldings number
$N$ is used, but one shall not falter since the applied process is
totally similar.

\item Without loss of generality, the $x_i$'s near the singularity
assume the form
    \begin{equation}
        \centering
        \label{x}
        x_i = a_i (t - t_c)^{p_i},
    \end{equation}
where $t_c$ is but an integration constant. Substituting (\ref{x})
in (\ref{sys}) and equating the exponents, one may find the
$p_i$'s that constitute the vector $\mathbf{p} = (p_1 , p_1 ,
\dots p_n)$. Having the exponents, we return to the system in
order to calculate the $a_i$'s and similarly form the vector
$\mathbf{a} = (a_1 , a_2 , \dots a_n)$. If $\mathbf{a}$ does
contain only real entries, it may give rise only to finite-time
singularities while if it has complex entries, it may give rise
only to non-finite-time singularities. It should be underlined
that $\mathbf{a}$ cannot assume zero entries. Taking that into
account, every set $(\mathbf{a}, \mathbf{p})$ is called a
\textit{dominant balance of the system}.

\item The next thing to do is calculate the Kovalevskaya matrix,
which is of the form
    \begin{equation}
        \centering
        \label{kov}
        R = \left(
\begin{matrix}
\frac{\partial f_1}{\partial x_1} & \frac{\partial f_1}{\partial x_2} & \cdots & \frac{\partial f_1}{\partial x_n}\\
\frac{\partial f_2}{\partial x_1} & \frac{\partial f_2}{\partial x_2} & \cdots & \frac{\partial f_2}{\partial x_n}\\
\vdots & \vdots & \ddots & \vdots\\
\frac{\partial f_n}{\partial x_1} & \frac{\partial f_n}{\partial x_2} & \cdots & \frac{\partial f_n}{\partial x_n}
\end{matrix}
\right) - \left(
\begin{matrix}
p_1 & 0 & \cdots & 0\\
0 & p_2 & \cdots & 0 \\
\vdots & \vdots & \ddots & \vdots\\
0 & 0 & \cdots & p_n
\end{matrix}
\right),
    \end{equation}
evaluate it in one of the dominant balances found in the previous
step and find its eigenvalues. The said eigenvalues have to be of
the form $(-1 , r_2 ,\dots r_n)$. If $r_2 , r_3 ,\dots r_n > 0$,
the singularity is general, that is independent of the initial
conditions. On the other hand, even if one of these eigenvalues is
negative, the singularity is local, that is dependent on the
initial conditions.
\end{itemize}

\section{Analysis of the Dynamical System via Dominant Balances}

We start with the presentation of  the relevant scalar field
dynamical system, just as presented in \cite{Wang:2018duq}, which
is characterized by the equations
\begin{equation}
    \centering
    \label{fr1}
    3 H^2 = \rho_\phi + \rho_{d m} + \rho_r,
\end{equation}
\begin{equation*}
    \centering
    \dot{\rho_\phi}+3 H(\rho_\phi + P_\phi) = - \rho_{d m} - 3 H \rho_{d m},
\end{equation*}
\begin{equation*}
    \centering
     \dot{\rho_\phi}+3 H(\rho_\phi + P_\phi) = - Q \rho_{d m}\dot{\phi},
\end{equation*}
\begin{equation*}
    \centering
    \rho_{d m} + 3 H \rho_{d m} = Q \rho_{d m}\dot{\phi},
\end{equation*}
where $\rho_\phi , \rho_{d m}, \rho_r$ are the energy densities of
the scalar field, dark matter and radiation, respectively,
$P_\phi$ is the pressure of the scalar field and $Q$ a constant
expressing the interaction between dark matter and dark energy.
Considering an inflationary potential of the form $V(\phi) \sim
e^{\lambda \phi}$ and introducing the dimensionless variables
\begin{equation}
    \centering
    \label{var}
    x_1 = \frac{\dot{\phi}}{\sqrt{6}H},
\end{equation}
\begin{equation*}
    \centering
    x_2 = \frac{\sqrt{V}}{\sqrt{3}H},
\end{equation*}
\begin{equation*}
    \centering
    x_3 = \frac{\sqrt{\rho_{d m}}}{\sqrt{3}H},
\end{equation*}
\begin{equation*}
    \centering
    x_4 = \frac{\sqrt{\rho_r}}{\sqrt{3}H},
\end{equation*}
eqs. (\ref{fr1}) become
\begin{equation}
    \centering
    \label{dyn}
    \frac{d x_1 }{d N} = - 3 x_1 - \frac{\sqrt{6}}{2}\lambda x_2^2 + \frac{1}{2}(3 x_1^3 - 3 x_1 x_2^2 - 3 x_1 x_3^2 + x_1 x_4^2 + 3 x_1) - \frac{\sqrt{6}}{2}Q(1 - x_1^2 - x_2^2 - x_3^2 - x_4^4),
\end{equation}
\begin{equation*}
    \centering
    \frac{d x_2 }{d N} = \frac{\sqrt{6}}{2}\lambda x_1 x_2 + \frac{1}{2}(3 x_1^2 x_2 - 3 x_2^3 - 3 x_2 x_3^2 + x_2 x_4^2 + 3 x_2),
\end{equation*}
\begin{equation*}
    \centering
    \frac{d x_3}{d N} = -\frac{3}{2}x_3 + \frac{1}{2}(3 x_1^2 x_3 - 3 x_2^2 x_3 - 3 x_3^3 + x_3 x_4^2 + 3 x_3),
\end{equation*}
\begin{equation*}
    \centering
    \frac{d x_4}{d N} = - 2 x_4 + \frac{1}{2}(3 x_1^2 x_4 - 3 x_2^2 x_4 - 3 x_3^2 x_4 + x_4^3 + 3 x_4),
\end{equation*}
with the Friedmann constraint  being $x_1^2 + x_2^2 + x_3^2 + x_4^2
= 1$ and $\lambda = |V'|/V$ a constant of order unity arising from
the Swampland (the prime denotes differentiation with respect to
the scalar field). Our next task, in order to apply the dominant
balances method, is to figure out all the possible distinct
mathematically consistent truncations of (\ref{dyn}); that is
vectors whose entries are dominant terms from each one of the
differential equations. It should be noted that since we are
working on a 4-manifold, constant and linear terms cannot be
dominant and are neglected whatsoever.

\subsection{1st mathematically consistent truncation}

The first mathematically consistent truncation of (\ref{dyn}) is
\begin{equation}
    \centering
    \label{tr1}
    \hat{f}_1 = \left(
    \begin{matrix}
    \frac{\sqrt{6}}{2}Q x_4^4\\
    -\frac{3}{2}x_2^3\\
    -\frac{3}{2}x_3^3\\
    \frac{1}{2}x_4^3
    \end{matrix}
    \right)
\end{equation}
and applying the method presented in the previous section, we easily find
\begin{equation}
    \centering
    \label{p1}
    \mathbf{p}=\left(-\frac{1}{3},-\frac{1}{2},-\frac{1}{2},-\frac{1}{2}\right)
\end{equation}
and the following 8 dominant balances:
\begin{equation}
    \centering
    \label{a1}
    \mathbf{a_1} = \left(-3\sqrt{\frac{3}{2}}Q, \frac{1}{\sqrt{3}},\frac{1}{\sqrt{3}},-i\right),
\end{equation}
\begin{equation*}
    \centering
    \mathbf{a_2} = \left(-3\sqrt{\frac{3}{2}}Q, \frac{1}{\sqrt{3}}, -\frac{1}{\sqrt{3}}, -i\right),
\end{equation*}
\begin{equation*}
    \centering
    \mathbf{a_3} = \left(-3\sqrt{\frac{3}{2}}Q, -\frac{1}{\sqrt{3}}, \frac{1}{\sqrt{3}}, -i\right),
\end{equation*}
\begin{equation*}
    \centering
    \mathbf{a_4} = \left(-3\sqrt{\frac{3}{2}}Q, -\frac{1}{\sqrt{3}}, -\frac{1}{\sqrt{3}}, -i\right),
\end{equation*}
\begin{equation*}
    \centering
    \mathbf{a_5} = \left(-3\sqrt{\frac{3}{2}}Q, \frac{1}{\sqrt{3}}, -\frac{1}{\sqrt{3}}, i\right),
\end{equation*}
\begin{equation*}
    \centering
    \mathbf{a_6} = \left(-3\sqrt{\frac{3}{2}}Q, \frac{1}{\sqrt{3}}, -\frac{1}{\sqrt{3}}, i\right),
\end{equation*}
\begin{equation*}
    \centering
    \mathbf{a_7} = \left(-3\sqrt{\frac{3}{2}}Q, -\frac{1}{\sqrt{3}}, \frac{1}{\sqrt{3}}, i\right),
\end{equation*}
\begin{equation*}
    \centering
    \mathbf{a_8} = \left(-3\sqrt{\frac{3}{2}}Q, -\frac{1}{\sqrt{3}}, -\frac{1}{\sqrt{3}}, i\right).
\end{equation*}
The vectors $\mathbf{a_i}$ have complex entries, so finite-time
singularities cannot occur. The Kovalevskaya matrix of (\ref{kov})
assumes the form,
\begin{equation}
    \centering
    \label{kov1}
    R = \left(
    \begin{matrix}
    \frac{1}{3} & 0 & 0 & 2 \sqrt{6}Q x_4^3\\
    0 & \frac{1}{2}-\frac{9}{2}x_2^2 & 0 & 0\\
    0 & 0 & \frac{1}{2}-\frac{9}{2}x_3^2 & 0\\
    0 & 0 & 0 & \frac{1}{2}+\frac{3}{2}x_4^2
    \end{matrix}
    \right).
\end{equation}
Substituting the dominant balances $(\mathbf{a_1}, \mathbf{p}),
(\mathbf{a_2}, \mathbf{p}), (\mathbf{a_3}, \mathbf{p})$ and
$(\mathbf{a_4}, \mathbf{p})$ in (\ref{kov1}), we obtain
\begin{equation}
    \centering
    \label{kov11}
    R = \left(
    \begin{matrix}
    \frac{1}{3} & 0 & 0 & 2 i \sqrt{6}Q\\
    0 & -1 & 0 & 0\\
    0 & 0 & -1 & 0\\
    0 & 0 & 0 & -1
    \end{matrix}
    \right),
\end{equation}
while for $(\mathbf{a_5}, \mathbf{p}), (\mathbf{a_6}, \mathbf{p}), (\mathbf{a_7}, \mathbf{p})$ and $(\mathbf{a_8}, \mathbf{p})$ in (\ref{kov1}), we get
\begin{equation}
    \centering
    \label{kov12}
    R = \left(
    \begin{matrix}
    \frac{1}{3} & 0 & 0 & -2 i \sqrt{6}Q\\
    0 & -1 & 0 & 0\\
    0 & 0 & -1 & 0\\
    0 & 0 & 0 & -1
    \end{matrix}
    \right).
\end{equation}
Both (\ref{kov11}) and (\ref{kov12}) have the same set of eigenvalues:
\begin{equation}
    \centering
    \label{r1}
    r = \left(-1, -1, -1, \frac{1}{3}\right)
\end{equation}
and, since $r_2 , r_3 < 0$, we conclude that only local singularities may occur.

\subsection{2nd mathematically consistent truncation}

The second mathematically consistent truncation of (\ref{dyn}) is
\begin{equation}
    \centering
    \label{tr2}
    \hat{f}_2 = \left(
    \begin{matrix}
    -\frac{3}{2} x_1 x_2^2\\
    \frac{\sqrt{6}}{2}\lambda x_1 x_2\\
    -\frac{3}{2}x_3^3\\
    \frac{1}{2}x_4^3
    \end{matrix}
    \right)
\end{equation}
and applying the aforementioned method we easily find
\begin{equation}
    \centering
    \label{p2}
    \mathbf{p}=\left(-1,-\frac{1}{2},-\frac{1}{2},-\frac{1}{2}\right)
\end{equation}
and the following 8 dominant balances:
\begin{equation}
    \centering
    \label{a2}
    \mathbf{a_1} = \left(-\frac{1}{\sqrt{6}\lambda}, \sqrt{\frac{2}{3}},\frac{1}{\sqrt{3}},-i\right),
\end{equation}
\begin{equation*}
    \centering
    \mathbf{a_2} = \left(-\frac{1}{\sqrt{6}\lambda}, \sqrt{\frac{2}{3}},\frac{1}{\sqrt{3}},i\right),
\end{equation*}
\begin{equation*}
    \centering
   \mathbf{a_3} = \left(-\frac{1}{\sqrt{6}\lambda}, \sqrt{\frac{2}{3}},-\frac{1}{\sqrt{3}},-i\right),
\end{equation*}
\begin{equation*}
    \centering
    \mathbf{a_4} = \left(-\frac{1}{\sqrt{6}\lambda}, \sqrt{\frac{2}{3}},-\frac{1}{\sqrt{3}},i\right),
\end{equation*}
\begin{equation*}
    \centering
    \mathbf{a_5} = \left(-\frac{1}{\sqrt{6}\lambda}, -\sqrt{\frac{2}{3}},\frac{1}{\sqrt{3}},-i\right),
\end{equation*}
\begin{equation*}
    \centering
    \mathbf{a_6} = \left(-\frac{1}{\sqrt{6}\lambda}, -\sqrt{\frac{2}{3}},\frac{1}{\sqrt{3}},i\right),
\end{equation*}
\begin{equation*}
    \centering
    \mathbf{a_7} = \left(-\frac{1}{\sqrt{6}\lambda}, -\sqrt{\frac{2}{3}},-\frac{1}{\sqrt{3}},-i\right),
\end{equation*}
\begin{equation*}
    \centering
    \mathbf{a_8} = \left(-\frac{1}{\sqrt{6}\lambda}, -\sqrt{\frac{2}{3}},-\frac{1}{\sqrt{3}},i\right).
\end{equation*}
Just like in the previous case, finite-time singularities cannot
occur. The Kovalevskaya matrix of (\ref{kov}) assumes the form
\begin{equation}
    \centering
    \label{kov2}
    R = \left(
    \begin{matrix}
    1-\frac{3}{2}x_2^2 & -3 x_1 x_2 & 0 & 0\\
    \sqrt{\frac{3}{2}}\lambda x_2 & \frac{1}{2}+\sqrt{\frac{3}{2}}\lambda x_2 & 0 & 0\\
    0 & 0 & \frac{1}{2}-\frac{9}{2}x_3^2 & 0\\
    0 & 0 & 0 & \frac{1}{2}+\frac{3}{2}x_4^2
    \end{matrix}
    \right).
\end{equation}
Substituting the dominant balances $(\mathbf{a_1}, \mathbf{p}),
(\mathbf{a_2}, \mathbf{p}), (\mathbf{a_3}, \mathbf{p})$ and
$(\mathbf{a_4}, \mathbf{p})$ in (\ref{kov2}), we obtain
\begin{equation}
    \centering
    \label{kov21}
    R = \left(
    \begin{matrix}
    0 & \frac{1}{\lambda} & 0 & 0\\
    \lambda & 0 & 0 & 0\\
    0 & 0 & -1 & 0\\
    0 & 0 & 0 & -1
    \end{matrix}
    \right),
\end{equation}
while for $(\mathbf{a_5}, \mathbf{p}), (\mathbf{a_6}, \mathbf{p}), (\mathbf{a_7}, \mathbf{p})$ and $(\mathbf{a_8}, \mathbf{p})$ in (\ref{kov2}), we obtain
\begin{equation}
    \centering
    \label{kov22}
    R = \left(
    \begin{matrix}
    0 & -\frac{1}{\lambda} & 0 & 0\\
    -\lambda & 0 & 0 & 0\\
    0 & 0 & -1 & 0\\
    0 & 0 & 0 & -1
    \end{matrix}
    \right).
\end{equation}

Both (\ref{kov21}) and (\ref{kov22}) have the same set of eigenvalues:
\begin{equation}
    \centering
    \label{r1}
    r = \left(-1, -1, -1, 1\right)
\end{equation}
and, because $r_2 , r_3 < 0$, non-general solutions are the only plausible scenario.

\subsection{3rd mathematically consistent truncation}

The third mathematically consistent truncation of (\ref{dyn}) is the following:
\begin{equation}
    \centering
    \label{tr3}
    \hat{f}_3=\left(
\begin{array}{c}
 \frac{\sqrt{6}}{2}Q  x_1^2 \\
 -\frac{3  x_2^3}{2} \\
 \frac{ x_3  x_4^2}{2} \\
 -\frac{3  x_3^2 x_4}{2}
\end{array}
\right).
\end{equation}
Using the dominant balances method, the following solution for the vector $\mathbf{p}$ is obtained:
\begin{equation}
    \centering
    \label{p}
    \mathbf{p}=\left(-1,-\frac{1}{2},-\frac{1}{2},-\frac{1}{2}\right).
\end{equation}
Accordingly, for $\mathbf{p}$ being the above, the following vector-solutions $\mathbf{a_i}$ are found:
\begin{equation}
    \centering
    \label{mathbf{a_1}}
    \mathbf{a_1}=\left(-\frac{2}{\sqrt{6}Q}, \frac{1}{\sqrt{3}},\frac{1}{\sqrt{3}},i\right),
\end{equation}
\begin{equation*}
    \centering
    \label{mathbf{a_2}}
   \mathbf{a_2}=\left(-\frac{2}{\sqrt{6}Q}, -\frac{1}{\sqrt{3}},-\frac{1}{\sqrt{3}},-i\right),
\end{equation*}
\begin{equation*}
    \centering
    \label{mathbf{a_3}}
    \mathbf{a_3}=\left(-\frac{2}{\sqrt{6}Q}, \frac{1}{\sqrt{3}},-\frac{1}{\sqrt{3}},-i\right),
\end{equation*}
\begin{equation*}
    \centering
    \label{math{a_4}}
   \mathbf{a_4}=\left(-\frac{2}{\sqrt{6}Q}, -\frac{1}{\sqrt{3}},\frac{1}{\sqrt{3}},i\right),
\end{equation*}
\begin{equation*}
    \centering
    \label{mathbf{a_5}}
    \mathbf{a_5}=\left(-\frac{2}{\sqrt{6}Q}, \frac{1}{\sqrt{3}},-\frac{1}{\sqrt{3}},i\right),
\end{equation*}
\begin{equation*}
    \centering
    \label{mathbf{a_6}}
    \mathbf{a_6}=\left(-\frac{2}{\sqrt{6}Q}, \frac{1}{\sqrt{3}},\frac{1}{\sqrt{3}},-i\right),
\end{equation*}
\begin{equation*}
    \centering
    \label{mathbf{a_7}}
    \mathbf{a_7}=\left(-\frac{2}{\sqrt{6}Q}, -\frac{1}{\sqrt{3}},\frac{1}{\sqrt{3}},-i\right),
\end{equation*}
\begin{equation*}
    \centering
    \label{mathbf{a_8}}
 \mathbf{a_8}=\left(-\frac{2}{\sqrt{6}Q}, -\frac{1}{\sqrt{3}},-\frac{1}{\sqrt{3}},i\right).
\end{equation*}
By calculating the Kovalevskaya matrix,
\begin{equation}
    \centering
    \label{kov3}
    R=\left(
\begin{array}{cccc}
 \sqrt{6}Q x_1+1 & 0 & 0 & 0 \\
 0 & -\frac{9 x_2^2}{2}+\frac{1}{2} & 0 & 0 \\
 0 & 0 & \frac{x_4^2}{2}+\frac{1}{2} & x_3x_4 \\
 0 & 0 & -3x_3x_4 & -\frac{3x_3^2}{2}+\frac{1}{2}
\end{array}
\right)
\end{equation}
we find its form for each $\mathbf{a_i}$:
\begin{equation}
    \centering
    \label{kov31}
    R(\mathbf{a_1})=R(\mathbf{a_2})=R(\mathbf{a_3})=R(\mathbf{a_4})=\left(
\begin{array}{cccc}
 -1 & 0 & 0 & 0 \\
 0 & -1 & 0 & 0 \\
 0 & 0 & 0 & \frac{i}{\sqrt{3}} \\
 0 & 0 & -\sqrt{3}i & 0
\end{array}
\right),
\end{equation}
\begin{equation*}
    \centering
    \label{kov32}
    R(\mathbf{a_5})=R(\mathbf{a_6})=R(\mathbf{a_7})=R(\mathbf{a_8})=\left(
\begin{array}{cccc}
 -1 & 0 & 0 & 0 \\
 0 & -1 & 0 & 0 \\
 0 & 0 & 0 & -\frac{i}{\sqrt{3}} \\
 0 & 0 & \sqrt{3}i & 0
\end{array}
\right).
\end{equation*}
Their eigenvalues are the same for all $\mathbf{a_i}$:
\begin{equation}
    \centering
    \label{eig3}
    (r_1, r_2, r_3, r_4)=(-1,-1,-1,1),
\end{equation}
and we see that $r_2 , r_3 < 0$. Hence, there is a limited set of initial conditions which leads the dynamical system to no finite-time singularities.

\subsection{4th mathematically consistent truncation}

The fourth mathematically consistent truncation of (\ref{dyn}) that can be obtained is
\begin{equation}
    \centering
    \label{tr4}
    \hat{f}_4=\left(
\begin{array}{c}
 \frac{\sqrt{6}}{2}Q  x_1^2 \\
 \frac{  x_2 x_4^2}{2} \\
 -\frac{3 x_2^2  x_3}{2} \\
 -\frac{3 x_3^2 x_4}{2}
\end{array}
\right).
\end{equation}
For (\ref{tr4}), the dominant balances $(\mathbf{a},\mathbf{p})$
are the same as for (\ref{tr3}). The corresponding Kovalevskaya
matrix is
\begin{equation}
    \centering
    \label{kov4}
    R=\left(
\begin{array}{cccc}
 \sqrt{6}Q  x_1+1 & 0 & 0 & 0 \\
 0 & \frac{ x_4^2}{2}+\frac{1}{2} & 0 &  x_2  x_4 \\
 0 & -3 x_2  x_3 & -\frac{3 x_2^2}{2}+\frac{1}{2} & 0 \\
 0 & 0 & -3 x_3  x_4 & -\frac{3 x_3^2}{2}+\frac{1}{2}
\end{array}
\right)
\end{equation}
and by substituting each dominant balance in (\ref{kov4}) we evaluate the corresponding eigenvalues:
\begin{equation}
    \centering
    \label{kov41}
    R(\mathbf{a_1})=R(\mathbf{a_2})=\left(
\begin{array}{cccc}
 -1 & 0 & 0 & 0 \\
 0 & 0 & 0 & \frac{i}{\sqrt{3}} \\
 0 & -1 & 0 & 0 \\
 0 & 0 & -\sqrt{3}i & 0
\end{array}
\right),
\end{equation}
\begin{equation*}
    \label{eig41}
    \centering
    (r_1, r_2, r_3, r_4)=(-1,-1,(-1)^{1/3}, -(-1)^{2/3}),
\end{equation*}
\begin{equation}
    \centering
    \label{kov42}
    R(\mathbf{a_3})=R(\mathbf{a_4})=\left(
\begin{array}{cccc}
 -1 & 0 & 0 & 0 \\
 0 & 0 & 0 & -\frac{i}{\sqrt{3}} \\
 0 & 1 & 0 & 0 \\
 0 & 0 & \sqrt{3}i & 0
\end{array}
\right),
\end{equation}
\begin{equation*}
    \centering
    \label{eig42}
    (r_1, r_2, r_3, r_4)=(-1,1,-(-1)^{1/3}, (-1)^{2/3}),
\end{equation*}
\begin{equation}
    \centering
    \label{kov43}
    R(\mathbf{a_5})=R(\mathbf{a_7})=\left(
\begin{array}{cccc}
 -1 & 0 & 0 & 0 \\
 0 & 0 & 0 & \frac{i}{\sqrt{3}} \\
 0 & 1 & 0 & 0 \\
 0 & 0 & -\sqrt{3}i & 0
\end{array}
\right),
\end{equation}
 \begin{equation*}
    \centering
    \label{eig43}
    (r_1, r_2, r_3, r_4)=(1,-(-1)^{1/3}, (-1)^{2/3}, -1/2),
\end{equation*}

\begin{equation}
    \centering
    \label{kov44}
    R(\mathbf{a_6})=R(\mathbf{a_8})=\left(
\begin{array}{cccc}
 -1 & 0 & 0 & 0 \\
 0 & 0 & 0 & -\frac{i}{\sqrt{3}} \\
 0 & -1 & 0 & 0 \\
 0 & 0 & -\sqrt{3}i & 0
\end{array}
\right),
\end{equation}
\begin{equation*}
    \centering
    \label{eig44}
    (r_1, r_2, r_3, r_4)=(-1,1,-(-1)^{1/3}, (-1)^{2/3}).
\end{equation*}
For $\mathbf{a_1}$, $\mathbf{a_2}$, $\mathbf{a_3}$,
$\mathbf{a_4}$, $\mathbf{a_6}$ and $\mathbf{a_8}$, (\ref{tr4})
leads to mathematically unappealing results, due to the complex
entries $(-1)^{1/3}$ and $(-1)^{2/3}$. Concerning $\mathbf{a_5}$
and $\mathbf{a_7}$,  the first Kovalevskaya eigenvalue is $r_1
\neq -1$ and therefore no conclusion can be reached.

\section{Conclusions}

In the present work we studied a Swampland related scalar field
theory in the presence of matter and radiation fluids, focusing on
the dynamical system that governs the cosmological system. Our aim
was to find whether finite-time singularities occur in the
dynamical system, and if yes, whether these correspond to general
or restricted sets of initial conditions. By employing the method
of dominant balances and its accompanying theorems we were able to
select the dominant balances of the dynamical system and
demonstrate that indeed finite-time singularities may actually
occur in the system, however these are limited types of
singularities, meaning that these occur for a very narrow range of
initial conditions. This is in contrast with the case that the
matter fluids were not present, thus the major effect of the
matter fluids is to essentially remove the finite-time
singularities from the phase space of the scalar field model.

{}

\end{document}